\documentstyle[aas2pp4]{article}

\newcommand{\MO}{M_{\sun}}   
\newcommand{\LO}{L_{\sun}}   

\slugcomment{accepted by The Astrophysical Journal}

\lefthead{Awaki et al.}
\righthead{X-ray emission from a merger remnant, NGC 7252}

\begin{document}

\title{X-ray emission from a merger remnant, NGC 7252,the ``Atoms-for-Peace'' galaxy }

\author{ Hisamitsu Awaki } 
\affil{Department of Physics, Faculty of Science, Ehime University} 
\authoraddr{Matsuyama, Ehime, 790-8577, Japan}

\authoremail{awaki@astro.phys.sci.ehime-u.ac.jp} 

\author{ Hironori Matsumoto\altaffilmark{1} } 

\affil{Department of Earth and Space Science, Osaka University,}
\authoraddr{1-1 Machikaneyama, Toyonaka, Osaka 560-0043, Japan} 

\and

\author{ Hiroshi Tomida }
\affil{Space Utilization Research Program, National Space Development Agency of Japan,}
\authoraddr{ 2-1-1, Sengen, Tsukuba, Ibaraki 305-8505, Japan }

\altaffiltext{1}{Center for Space Research, Massachusetts Institute of Technology, 77 Massachusetts Avenue, NE80-6045, Cambridge, MA02139-4307, USA }

\begin{abstract}
We observed a nearby merger remnant NGC 7252 with the X-ray satellite 
{\it ASCA}, and detected X-ray emission with the X-ray flux of 
(1.8$\pm$0.3)$\times10^{-13}$ ergs s$^{-1}$ cm$^{-2}$ in the 0.5--10 keV 
band. This corresponds to the X-ray luminosity of 8.1$\times10^{40}$ ergs 
s$^{-1}$. The X-ray emission is well described with a two-component model: 
a soft component with $kT = 0.72\pm$0.13 keV and a hard component with 
$kT > 5.1$ keV. Although NGC 7252 is referred to as a dynamically young 
protoelliptical, the 0.5--4 keV luminosity of the soft component is about 
2$\times10^{40}$ ergs s$^{-1}$, which is low for an early-type 
galaxy. The ratio of $L_{\rm{X}}/L_{\rm{FIR}}$ suggests that the soft 
component originated from the hot gas due to star formation. Its low 
luminosity can be explained by the gas ejection from the galaxy as galaxy 
winds.
Our observation reveals the existence of hard X-ray emission with the 
2--10 keV luminosity of 5.6$\times10^{40}$ ergs s$^{-1}$. This may indicate 
the existence of nuclear activity or intermediate-mass black hole in NGC 7252.
\end{abstract}


\section{Introduction}
NGC 7252 is a nearby galaxy (63.2 Mpc) with tidal tails in the optical band.
The tidal tails are thought to be the result of a galaxy merger of two 
spiral galaxies (e.g. Mihos, Dubinski \& Hernquist 1998).  The merger 
phenomenon probably induced a violent starformation in the whole body 
(e.g. Hibbard et al. 1994, hereafter H94).  We can observe traces of the 
violent starformation which occurred less than 1 Gyr ago; for instance, the 
presence of an ``E+A'' spectrum (Schweitzer 1982), the absence 
of H I gas in the inner region (Hibbard \& van Gorkom 1996), and the
existence of young globular clusters (e.g. Whitemore et al. 1997). 
In the central region, its surface profile is well represented by the 
$r^{1/4}$ law which is seen in early-type galaxies (e.g. Schweizer 1982, 
Stanford \& Bushouse 1991). 
Numerical simulations predict that the tidal tails will 
disappear within a few Gyr, and then this galaxy will be recognized as an 
early-type galaxy (e.g. Hibbard \& Mihos 1995). Therefore, this galaxy is also 
called ``a dynamically young protoelliptical''.

Early-type galaxies sometimes have powerful X-ray emission which originates 
from a hot gaseous halo surrounding the 
galaxies. Their X-ray luminosities range from 10$^{39}$ to 10$^{42}$ ergs 
s$^{-1}$ (e.g. Forman, Jones, \& Tucker 1985; Canizares, Fabbiano, \& 
Trinchieri 1987). The total amount of the hot gaseous halo and the metal 
abundance are estimated to be 10$^{8}$--10$^{11} \MO$ and 
sub-solar, respectively (e.g. Awaki et al. 1994; 
Matsushita, Ohashi, \& Makishima 2000). 
 In order to reveal the evolution of the hot gas, many observational and 
theoretical works have been performed. Since NGC 7252 is evolving into an 
early-type galaxy, this galaxy is one of the best 
objects for investigating the evolution of the hot gas surrounding early-type 
galaxies. H94 detected X-rays from NGC 7252 with the {\it ROSAT} PSPC and 
HRI, and found that the X-ray emitting region is marginally extended, which 
indicates the existence of hot gas in NGC 7252.  However, they could not 
determine its accurate spectral shape. Using the {\it ASCA}'s wide-band 
spectroscopy, we reveal the nature of the hot gas in NGC 7252. 

\section{Observation and Results}
NGC 7252 was observed with {\it ASCA} on 1998 November 16--18 (Tanaka, 
Inoue, \& Holt 1994). Since the intense shower of Leonid meteors was expected 
to appear on November 17, the high voltage of the GIS was turned off in 
order to minimize the impact to the GIS. Therefore, the galaxy was observed 
with only the solid-state imaging spectrometer (SIS). The SIS consists of two 
detectors, called SIS0 and SIS1. The SIS data were taken in the 1 CCD FAINT 
mode at all bit-rates. The data reduction was performed by the {\it ASCA} 
Guest Observatory at NASA/GSFC using the following main criteria: (1) the 
angle from bright Earth's limb $> 20\arcdeg$, (2) the elevation angle from 
Earth's limb $> 10\arcdeg$, (3) the geometrical cut off rigidity $> 6$ GeV 
c$^{-1}$.  The detail of the selection criteria is provided from the 
{\it ASCA} Guest Observatory. The on-source exposure time after the above 
screening is about 43.5 ks.

We obtained an X-ray image in the 0.5--10 keV band with the data from SIS0 
and SIS1.  Figure 1 shows the X-ray contour map overlaid on the digitized sky
survey map of NGC 7252. We accumulated SIS0 and SIS1 data within 
a 6$\arcmin$ diameter circle centered on the X-ray source; a background
spectrum is taken from a non-source region $> 8\arcmin$ in diameter. 
We combine energy bins to contain more than 50 counts in each bin and
apply a $\chi^2$ method in the XSPEC 11.00 package for spectral fitting.
The spectrum after background subtraction is shown in Figure 2.

We first attempted to fit a single-component model, a thermal model or
a power law model, to the spectrum. However, the spectrum was 
characterized 
with $kT\sim$2.3 keV or $\it{\Gamma}\sim$2.3 with $\chi_{\nu}^{2} > 1.7$.  
Next, we introduced a two-component model, thin thermal plus thermal 
bremsstrahlung model. This model provided an acceptable fit to the data (
$\chi_{\nu}^{2} \sim 1.2$), with the absorption column fixed at the Galactic
value of 2$\times10^{20}$ cm$^{-2}$ (Stark et al. 1990). 
Since the confidence region of the metal abundance $A_{\rm{RS}}$ was too
large, we fixed $A_{\rm{RS}}$ at one solar metal abundance (Table 1).
The thin thermal and the thermal bremsstrahlung models are 
characterized with $kT = 0.72\pm$0.13 keV and $kT > 5.1$ keV, respectively.
The best-fit model is shown in Figure 2. Using this model, the X-ray flux is 
(1.8$\pm$0.3)$\times10^{-13}$ ergs s$^{-1}$ cm$^{-2}$ in the 0.5--10 keV 
band, which corresponds to the X-ray luminosity of 8.1$\times10^{40}$ ergs 
s$^{-1}$ assuming the distance of 63.2 Mpc. The X-ray flux in the {\it ROSAT}
band (0.2--4 keV) is about 1.0$\times10^{-13}$ ergs s$^{-1}$ cm$^{-2}$, which 
is consistent with that obtained by H94, although they did not
determine the spectral shape of the X-rays from NGC 7252, and, hence assumed
 a thin thermal emission with $kT = 1$ keV and metal abundance of 1 solar.

\section{Discussion}
\subsection{Origin of the Soft Component}
The {\it ROSAT} observation by H94 found a soft X-ray emission which is 
marginally extended, which indicates the existence of the hot gas in NGC 
7252. Our {\it ASCA} observation finds that the soft X-ray emission has the 
characteristic temperature of $kT = 0.72\pm$0.13 keV, and the observed X-ray 
flux in the 0.5--4 keV band is about 4$\times10^{-14}$ ergs s$^{-1}$ 
cm$^{-2}$, which corresponds to the X-ray luminosity of the 
2$\times10^{40}$ ergs s$^{-1}$ after the correction of the Galactic absorption.
As compared with X-ray properties of early type galaxies, the soft X-ray 
luminosity of NGC 7252 is low for an early-type galaxy. Our result is 
consistent with 
a claim by Fabbiano \& Schweizer (1995), who pointed out that NGC 7252 
belongs to group 1 in the $L_{\rm{B}}-L_{\rm{X}}$ diagram of E and S0 
galaxies (see figure 7 in their paper). In group 1, the contribution of the 
soft component to the total X-ray emission is considered to be small
(e.g. Fabbiano \& Schweizer 1995, Matsumoto et al. 1997).   

NGC 7252 is a post-starburst galaxy as well as a dynamically young 
protoelliptical (e.g. Schweizer 1982, Stanford \& Bushouse 1991). Both types 
of galaxies sometimes have hot gas surrounding the galaxy. We took the 
luminosity ratio of $L_{\rm{X}}/L_{\rm{FIR}}$ to find a possible origin of 
the hot gas in the 
galaxy. Figure 3 shows the correlation plot of the X-ray luminosities in the 
0.5--4 keV band and the far-infrared luminosities. 
The region between two solid lines in Figure 3 indicates the luminosity ratio 
of about 10$^{-4}$ which most starburst/H II galaxies have 
( David, Jones, \& Forman (1992)). The open circles in Figure 3 show the data
for early-type galaxies. Their X-ray data are taken from Canizares et al. 
(1987), and their infrared data are deduced from IRAS 60 and 100 $\micron$ 
fluxes (taken from Knapp et al. (1989)) using the usual ``bolometric'' 
far-infrared brightness (e.g. Devereux \& Eales 1989)
\begin{equation}
  F_{\rm{FIR}} = 1.24 \times 10^{-14} (2.58F_{\rm{60\micron}} + F_{\rm{100\micron}})\rm{~erg~cm^{-2}~s^{-1}}\;,
\end{equation}
where the 60 and 100 $\micron$ flux densities are expressed in mJy. 
The far-infrared light from early-type galaxies is emitted from cool 
interstellar matter, and thus their luminosities are smaller than those of 
starburst galaxies (Knapp et al. 1989; Bregman et al. 1998).  Therefore, 
early-type galaxies lie in the upper left side in the figure. 
 NGC 7252 is a luminous far-infrared object with 
$L_{\rm{FIR}} \sim 4\times10^{10} \LO$ (Dupraz et al. 1990). 
We find that the luminosity ratio of $L_{\rm{X}}/L_{\rm{FIR}}$ of NGC 7252
is about 10$^{-4}$, and find that the position of a closed circle in 
Figure 3 presenting NGC 7252 is within the region of starburst galaxies. 
The luminosity ratio suggests that the soft X-ray emission is associated 
with the starburst activity. Note that the X-ray luminosity of NGC 7252 
includes X-ray emissions from both soft ($L_{\rm{X}}=2\times10^{40}$) and 
hard ($L_{\rm{X}}=2.7\times10^{40}$) components, because Canizares et al. 
(1987) and David et al. (1992) did not distinguish soft and hard components 
when they estimated the X-ray luminosities. 
   
In order to estimate the total amount of the hot gas and the total energy, 
we obtained the volume emission measure, $\varepsilon=\int n^{2} dV$, of 
$(6.0_{-2.0}^{+3.1})\times10^{62}$ cm$^{-3}$ from the spectral fit of the 
X-ray spectrum with solar abundance ($A_{\rm{RS}}$ = 1.0).
This emission measure is the lowest compared to the sample of early-type 
galaxies by Matsumoto et al. (1997).  We note that the sample by 
Matsumoto et al. (1997) have a mean abundance of 0.3, and that the volume 
emission measure of a thin thermal plasma is a strong function of the metal 
abundance.  We obtained 
the emission measure $\varepsilon \sim 1.9 \times10^{63}$ cm$^{-3}$ in 
$A_{\rm{RS}}$=0.3, and confirmed that the emission measure is still 
low among their sample.
Read \& Ponman (1998) found that the {\it ROSAT} PSPC source appeared  
centrally concentrated and compact, though a couple of tentative extended 
features were seen. The X-ray image by the {\it ROSAT} observation is 
consistent with the H${\rm{\alpha}}$ image reported by H94. In starburst 
galaxies, the X-ray emitting region is well correlated with the 
H${\rm{\alpha}}$
emitting region (for example, NGC 1808 (Dahlem, Hartner, \& Junkes 1994)). 
Therefore, we assumed that the hot gas fills the inner region of $r$=2 kpc 
with a filling factor $f$, where $r$ is a radius of the hot plasma, because 
H94 pointed out that about 90 \% of the H${\rm{\alpha}}$ line emission is in 
$r < 7^{\prime\prime}$ ($\sim$ 2.5 kpc at D = 63.2 Mpc). We obtain the 
mean electron number density to be $<n_{e}> \sim 2\times 10^{-2} 
(r/2 \rm{kpc})^{-3/2} \it{f}^{-1/2}$ cm$^{-3}$. The total mass and the 
total energy are estimated to be 2$\times10^{7} (r/2 \rm{kpc})^{3/2} 
\it{f}^{1/2} \MO$ and 7$\times 10^{55} (r/2 \rm{kpc})^{3/2} \it{f}^{1/2}$ 
ergs, respectively. They indicate that in 
starburst activity, NGC 7252 can stand comparison with nearby active 
starburst galaxies (e.g. M82 and NGC 253 (Fabbiano 1988) ). 

Ciotti et al. (1991) classified the early-type galaxies into three 
evolutionary 
stages: wind, outflow, and inflow phases. Since it is thought that the 
intense star formation occurred within the last 1 Gyr (Schweitzer 1982; 
Hibbard \& van Gorkom 1996) and numerical simulations have suggested that
NGC 7252 evolves to an early-type galaxy, we consider that NGC 7252 belongs 
to the wind phase, in which the thermalized gas due to collisions between 
stellar winds and/or between winds and ambient gas is ejected from the 
galaxy.  The low X-ray luminosity of NGC 7252 can be explained by the gas 
ejection from the galaxy. Based on the evolutional model
by Ciotti et al. (1991), within a few Gyr, NGC 7252 will change to outflow 
phase, and then NGC 7252 will be surrounded by a large amount of hot gas.

\subsection{Hard Component}
Read \& Ponman (1998) suggested the existence of hard X-ray sources in 
NGC 7252, since the fitted absorbing column appears larger than the Galactic 
value. The {\it ASCA}'s wide-band spectroscopy resolved the hard X-ray 
emission with the X-ray luminosity of about 5.6$\times10^{40}$ ergs 
s$^{-1}$ in the 2--10 keV band. The spectrum is characterized by the 
thermal model with $kT > 5.1$ keV or the power law model with photon index 
of 1.2. 

The {\it ROSAT} HRI found two bright X-ray sources (Read \& Ponman 1998). 
One is located within 7$^{\prime\prime}$ from the center of NGC 7252, and 
its X-ray luminosity is estimated to be 3$\times10^{40}$ ergs s$^{-1}$ 
assuming the thermal emission with $kT = 3$ keV.  The other is located in 
the eastern extension seen in the PSPC, and its X-ray luminosity is 
estimated to be 1.6$\times10^{40}$ ergs s$^{-1}$. {\it ASCA} cannot separate 
these two sources due to its poor spatial resolution. 
Since the X-ray luminosity observed by {\it ASCA} is nearly equal to the sum 
of the X-ray luminosities of the two sources, we consider that most of the 
hard X-ray emission come from these two sources.
From the brightness of these two sources in the {\it ROSAT} HRI band, the 
significant fraction of the hard X-rays detected with {\it ASCA} is probably 
emitted from the former bright HRI source. 
Our result indicates the existence of a hard X-ray source with
$L_{\rm{X}} \sim 10^{40}$ erg s$^{-1}$ in the central region. The hard X-ray 
source may be a low-luminosity AGN or an intermediate-mass black hole
(10$^{3-6} \MO$) (Colbert \& Mushotzky 1999; Matsumoto et al. 2001; 
Kaaret et al. 2001). 
 
\section{Conclusion}

We analyzed {\it ASCA} data for a merging galaxy NGC 7252. The X-ray spectrum 
is well described by a two-component model, soft plus hard components. The
soft component is characterized by a thin thermal plasma model with 
$kT\sim$0.73keV. From the $L_{\rm{X}}/L_{\rm{FIR}}$, the thin thermal emission
probably originates from the hot gas due to star formation. Assuming the 
size of hot gas of 2 kpc, the total amount of hot gas is estimated to be
2$\times10^7 f^{1/2} \MO$, which indicate that NGC 7252 still have large 
starburst activity comparable to those of nearby active starburst galaxies.
 Numerical simulations and observational results suggest
that this galaxy is in the post-starburst phase. Therefore, we can explain the
low X-ray luminosity of NGC 7252 due to ejection of hot gas from the galaxy. 
Based on the evolutional model by Ciotti et al. (1990), this galaxy will be 
surrounded by a large amount of hot gas within a few Gyr.

We resolve the hard X-ray emission characterized by the thermal emission 
with $kT > 5.1$ keV or a power law emission with $\it{\Gamma}\sim$1.2 using
the {\it ASCA}'s wide-band spectroscopy. The {\it ROSAT} HRI found two bright 
point sources in this galaxy.  These sources are possible objects for the 
hard X-ray emission.  Our {\it ASCA} and the {\it ROSAT} results 
may suggest the existence of a low-luminosity AGN or an intermediate-mass 
black hole.

This galaxy is one of the best targets to investigate the evolution of 
merging galaxies and the connection between starburst galaxies and AGNs. The 
{\it XMM-Newton}, {\it Chandra}, and future missions will solve the current 
problems.

\acknowledgements
The authors thank all members of the {\it ASCA} team, and the staff in 
the {\it ASCA} GOF for providing {\it ASCA} data.  We thank  Ms. D. Gage 
and Ms. M. Awaki for their help. We thank the staff in the SkyView data 
surveys for providing the digitized sky survey image of NGC 7252. This 
research is supported by JSPS (HM), and  is carried out as part of the 
``Ground Research Announcement for Space Utilization'' promoted by the Japan 
Space Forum (HA).

\clearpage
\begin{deluxetable}{lcccccc}
\footnotesize
\tablecaption{Result of Spectral Fit \label{tbl_1}}
\tablewidth{0pt}
\tablehead{
\colhead{Model$^{a}$} & \colhead{$N_{\rm{H}}^{b}$} & \colhead{$kT_{\rm{RS}}$} &
\colhead{$A_{\rm{RS}}$} &  \colhead{$kT_{\rm{BR}}/\it{\Gamma}$} & \colhead{$\chi_{\nu}^{2}$/d.o.f.} 
\nl
   & ($\times10^{22}$ cm$^{-2}$) & (keV) &  (solar) &  (keV) & \nl
}
\startdata
RS + BR  & 0.02(fixed) & 0.73 (0.60-0.86) & 5.0 (0.05$<$ ) & 4.5 $<$ &  1.24/13 \nl
RS + BR  & 0.02(fixed) & 0.72 (0.61-0.85) & 1.0 (fixed)    & 5.1 $<$ &  1.16/14 \nl
\nl
RS + PL & 0.02(fixed) & 0.73 (0.60-0.86) & 5.0 (0.03$<$) & 1.21(0.45-1.88) & 1.25/13 \nl
RS + PL & 0.02(fixed) & 0.73 (0.60-0.86) & 1.0 (fixed) & 1.21(0.72-1.86) & 1.16/14 \nl
\tablenotetext{a}{the RS + BR indicates a Raymond-Smith optically thin plasma (RS) plus a bremsstrahlung (BR) model, and the RS + PL indicates a RS plus a power law (PL) model.}
\tablenotetext{b}{the column density is fixed at the Galactic absorbing column density given in Stark et al. (1992).}
\tablecomments{errors are 90 \% confidence level.}
\enddata
\end{deluxetable}

\newpage
\onecolumn
\begin{figure}
\epsscale{0.7}
\plotone{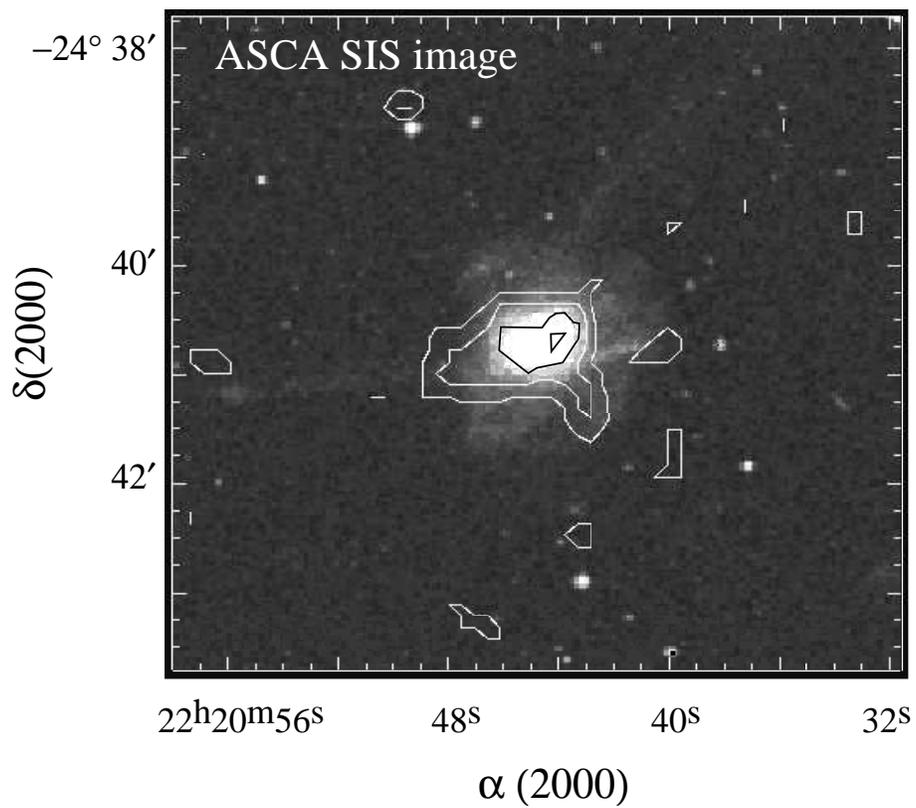}
\caption[f1.eps]{{\it ASCA} SIS contour map overlaid the digitized sky 
survey image of NGC 7252. The map has been smoothed with $\sigma$=30$\arcsec$.
 The contours are at 2, 3, 4, and 5 $\sigma$ fluctuation levels of the 
SIS background. 
\label{fig1}}
\end{figure}

\begin{figure}
\epsscale{0.6}
\plotone{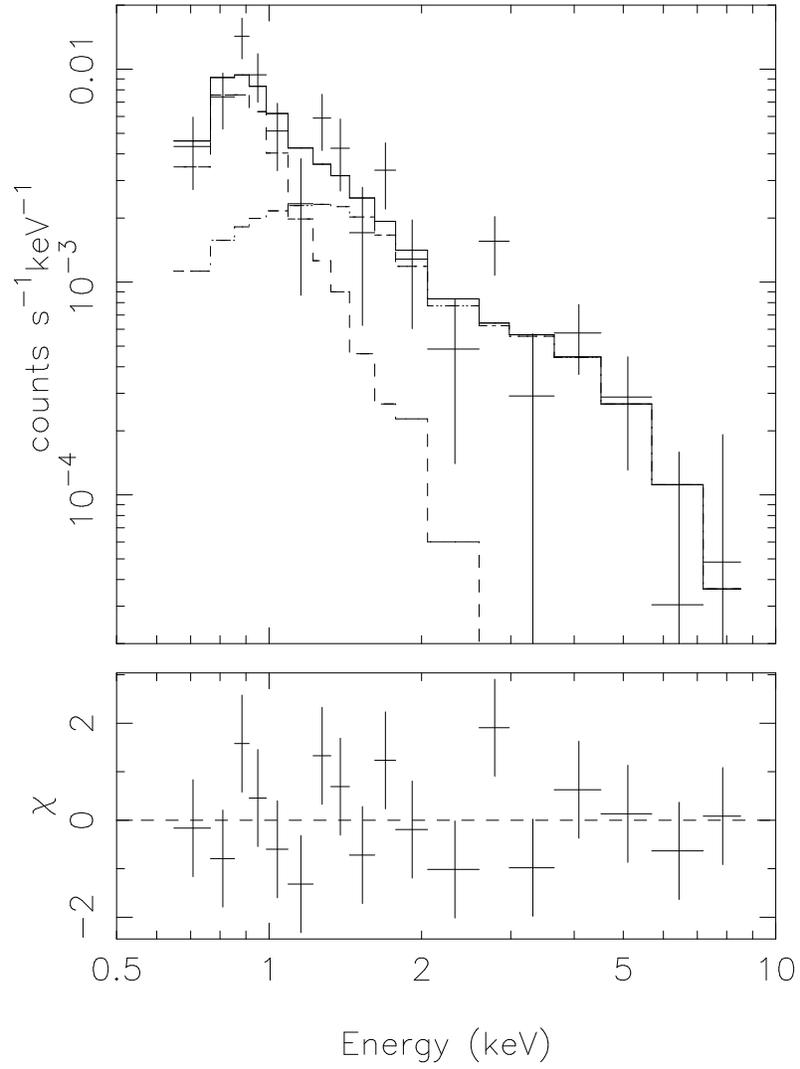}
\caption[f2.eps]{{\it ASCA} SIS spectrum of NGC 7252 fitted with a two-component model, consisting of a soft RS plasma component (dashed line) and a hard bremsstrahlung
component (dot-dashed line).  
\label{fig2}}
\end{figure}

\begin{figure}
\epsscale{0.75}
\plotone{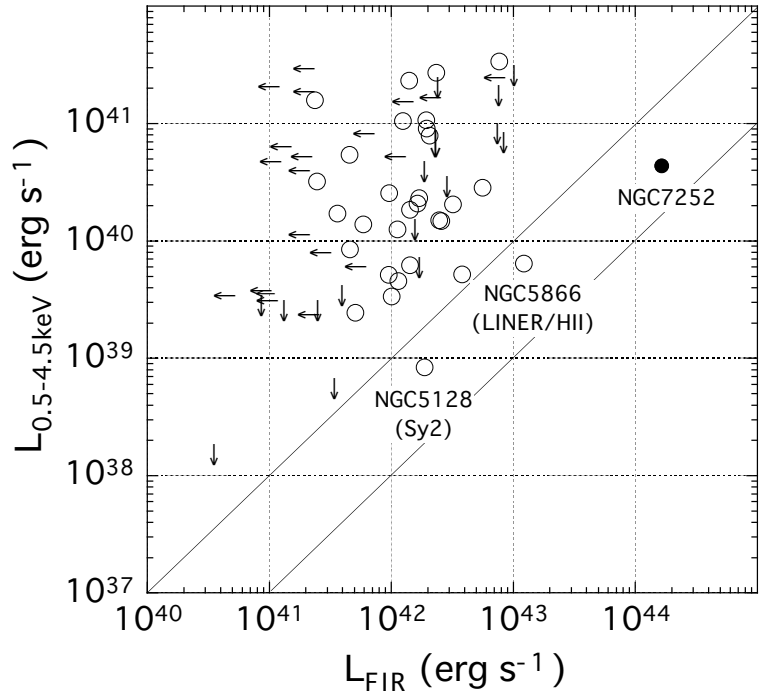}
\caption[f3.eps]{Correlation plot between far-infrared and the 0.5--4.5 
keV band luminosities. Open circles and one closed circle show the data for 
early-type galaxies and for NGC 7252, respectively. The region between
two solid lines indicates the region of spiral galaxies found by David
et al. (1991). A couple of early-type galaxies locate in the region of spiral
galaxies, but they are also classified as Seyfert/LINER. 
\label{fig3}}
\end{figure}

\end{document}